\documentclass{aa}  
\usepackage{graphicx}
\usepackage{txfonts}
\begin{document}
   \title{Multiple protostellar systems}

   \subtitle{II. A high resolution near-infrared imaging survey in
   nearby star-forming regions\thanks{Based on observations collected
   at the European Southern Observatory, Chile, under programs
   71.C-0716, 072.C-0321, 073.C-0450 and 075.C-0312.}}

   \author{G. Duch\^ene\inst{1}\thanks{\emph{Present
          address:} Department of Astronomy, 601 Campbell Hall,
          University of California Berkeley, Berkeley, CA 94720-3411, USA}
          \and
	  S. Bontemps\inst{2}
	  \and
          J. Bouvier\inst{1}
	  \and
	  P. Andr\'e\inst{3,4}
	  \and
	  A. A. Djupvik\inst{5}
	  \and
	  A. M. Ghez\inst{6}
          }


   \institute{Laboratoire d'Astrophysique de Grenoble, BP 53, 38041
              Grenoble cedex 9, France \\
              \email{gaspard.duchene@obs.ujf-grenoble.fr}
         \and
   OASU/LAB-UMR5804, CNRS, Universit\'e Bordeaux I, 2 rue de
              l'Observatoire, BP 89, 33270 Floirac, France
   \and 
   CEA/DSM/DAPNIA, Service d'Astrophysique, CEA Saclay, 91191
              Gif-sur-Yvette, France
   \and
   Laboratoire AIM, Unit\'e Mixte de Recherche CEA/DSM -- CNRS --
              Universit\'e Paris Diderot, C.E.A. Saclay, France
   \and
   Nordic Optical Telescope, Apdo 474, 38700 Santa Cruz de La Palma,
              Spain
   \and
   Department of Physics and Astronomy, UCLA, Los Angeles, CA
              90095-1562, USA}

   \date{Received 12/02/2007 ; accepted 03/10/2007}

 
  \abstract 
  {Multiple systems are the product of protostellar core
  fragmentation. Studying their statistical properties in young
  stellar populations therefore probes the physical processes at play
  during star formation.}
  {Our project endeavors to obtain a robust
  view of multiplicity among embedded Class\,I and Flat Spectrum
  protostars in a wide array of nearby molecular clouds to disentangle
  ``universal'' from cloud-dependent processes.}  
  {We have used near-infrared adaptive optics observations at the VLT
  through the $H$, $K_s$ and $L'$ filters to search for tight
  companions to 45 Class\,I and Flat Spectrum protostars located in 4
  different molecular clouds (Taurus-Auriga, Ophiuchus, Serpens and
  L1641 in Orion). We complemented these observations with published
  high-resolution surveys of 13 additional objects in Taurus and
  Ophiuchus.}
  {We found multiplicity
  rates of 32$\pm$6\% and 47$\pm$8\% over the 45--1400\,AU and
  14--1400\,AU separation ranges, respectively. These rates are in
  excellent agreement with those previously found among T\,Tauri stars
  in Taurus and Ophiuchus, and represent an excess of a factor
  $\sim$1.7 over the multiplicity rate of solar-type field stars. We
  found no non-hierarchical triple systems, nor any quadruple or
  higher-order systems. No significant cloud-to-cloud difference has
  been found, except for the fact that all companions to low-mass
  Orion protostars are found within 100\,AU of their primaries whereas
  companions found in other clouds span the whole range probed here.}
  {Based on this survey, we conclude that core fragmentation always
  yields a high initial multiplicity rate, even in giant molecular
  clouds such as the Orion cloud or in clustered stellar populations
  as in Serpens, in contrast with predictions of numerical
  simulations. The lower multiplicity rate observed in clustered
  Class\,II and Class\,III populations can be accounted for by a
  universal set of properties for young systems and subsequent
  ejections through close encounters with unrelated cluster members.}

  \maketitle


\section{Introduction}

The prevalence of binary and multiple systems among stellar
populations in our Galaxy is generally understood as a consequence of
the natural tendency of prestellar cores for fragmentation during or
immediately after their free-fall collapse (see Tohline 2002 for a
detailed review). Numerical simulations have long predicted that this
fragmentation, hence the frequency and properties of multiple systems,
is strongly dependent on the initial conditions reigning in the core
(e.g., Bonnell et al. 1992; Durisen \& Sterzik 1994; Boss 2002;
Goodwin et al. 2004b). Millimeter observations over the last two
decades have provided a detailed view of the initial conditions of
star formation. For instance, the basic properties of individual
prestellar cores, outer radius and central density, differ
significantly from one molecular cloud to another (Motte \& Andr\'e
2001). It is also likely that the temperature in prestellar cores vary
from cloud to cloud depending on the strength of the interstellar
radiation field (e.g., Stamatellos et al. 2007). Finally, there is now
good evidence that protostellar collapse is generally more violent in
a cluster-forming environment than in isolated dense cores, e.g.,
induced by strong external disturbances as opposed to spontaneous or
self-initiated (Andr\'e et al. 2003; Belloche et al. 2002,
2006). Considering these differences, one may therefore expect to
observe substantial differences in the properties of multiple systems
in independent stellar populations.

Early high-angular resolution surveys of Myr-old T\,Tauri stars in
nearby T associations, such as Taurus and Ophiuchus (Ghez et al. 1993;
Leinert et al. 1993; Reipurth \& Zinnecker 1993) revealed a
significant excess over the multiplicity rate among field stars
(Duquennoy \& Mayor 1991, hereafter DM91; Fischer \& Marcy 1992). It
was then rapidly discovered that clustered populations of equally
young stars, such as the Orion Nebula Cluster (ONC), do not possess
such a high multiplicity rate and rather resemble field stars from
this point of view (Padgett et al. 1997; Petr et al. 1998; Duch\^ene
et al. 1999). While this could be evidence that supports the
environment-dependent fragmentation of prestellar cores, alternative
explanations cannot be excluded. In particular, dynamical interactions
with unrelated cloud members can disrupt most wide companions in less
than 1\,Myr in the densest clusters (Kroupa 1995). In other words,
T\,Tauri multiple systems have had sufficient time to evolve since
their formation, so that their observed properties may not be
considered as directly representative of core fragmentation. To
circumvent this problem, observations of less evolved, embedded, young
stellar objects (YSOs) are required in order to determine the pristine
properties of multiple systems.

Radio interferometric studies of embedded Class\,0 and Class\,I
sources have hinted at a high overall multiplicity rate, comparable to
that observed among T\,Tauri stars in non-clustered populations
(Looney et al. 2000; Reipurth et al. 2002; 2004). Non uniform and
limited samples, as well as the fact that not all YSOs emits strongly
at centimeter wavelengths, prevent conclusive comparisons at this
point, however. In parallel to this effort, Haisch et al. (2002, 2004)
and Duch\^ene et al. (2004, hereafter D04) conducted direct
near-infrared imaging surveys of Class\,I protostars. These surveys
also found a high multiplicity rate, consistent with that of somewhat
more evolved T\,Tauri stars in the same clouds. D04 found marginal
evidence that the multiplicity rate of these sources decreases on a
timescale of $\sim10^5$\,yr, possibly a result of the internal or
external dynamical ejection of wide companions. No strong evidence was
found for a variation of the multiplicity rate of Class\,I sources
from one cloud to another, in part because of small sample sizes and
of the limited range of projected separations probed by these
surveys. Whether fragmentation does indeed depend on environmental
conditions remains to date a theoretical/numerical predictions that
has not yet been confirmed observationally. In particular, the outcome
of core fragmentation in a very rich molecular cloud such as Orion
remains unknown.

To extend the analysis of D04, we have undertaken a high-angular
resolution survey of embedded Class\,I protostars sampling the Serpens
and Orion molecular clouds in addition to Taurus and Ophiuchus. This
paper is organized as follows: we present our sample and observations
in Section\,\ref{sect:obs} and the results in
Section\,\ref{sect:results}. In Section\,\ref{sect:discus}, we analyze
these results in view of other multiplicity surveys and of predictions
of star formation theories. We summarize our findings in
Section\,\ref{sect:concl}.


\section{Sample, observations and data reduction}
\label{sect:obs}


\subsection{Sample definition}

Our objective is twofold in this survey: to extend the range of
spatial scales over which companions are searched for and to obtain a
valuable comparison between star-forming regions that represent
different types of star formation. We therefore performed our survey
in the Taurus-Auriga, Ophiuchus, Serpens and Orion molecular
clouds. More precisely, our survey in Orion focused on L1641, one of
the clouds that is currently most actively forming stars. Throughout
this paper, we adopt distances of 140\,pc to the Taurus and Ophiuchus
clouds, 260\,pc to the Serpens cloud and 450\,pc to L1641 (Bertout et
al. 1999; Bertout \& Genova 2006; Bontemps et al. 2001; Festin 1998;
Warren \& Hesser 1977).

\begin{table*}
\caption{\label{tab:sample_obs}Sample and observation log.}
\centering
\begin{tabular}{|ll|ccccc|c|cccl|}
\hline
 Target & Alt. Name & $K$ & $\alpha_{\mathrm IR}$ &
 Class &  $L_{\mathrm bol} (L_\odot)$ & Ref.$^{\mathrm{a}}$ &
 Mult.$^{\mathrm{b}}$ & Filters & $SR_{L'}$ & $SNR_{L'}$ &
 Obs. Date$^{\mathrm{c}}$ (UT) \\
\hline
\multicolumn{12}{|c|}{Taurus}\\
\hline
04113+2758 & & 7.79 & -0.01 & FS & $>$1.6 & 4,7 & B & $HK_sL'$ & 0.54
 & 8000 & 03/12/03 \\
04239+2436 & & 10.58 & 1.27 & I & 1.3 & 4,7 & B$^\star$ & $HK_sL'$ &
0.27 & 2000 & 04/12/03 \\
04263+2426 & Haro 6-10 & 7.80 & 1.02 & I & 0.74 & 4,7 & B & $HK_sL'$ &
 $\sim0.70$ & 8000 & 03/12/03 \\
04287+1801$^{\mathrm{d}}$ & L1551 IRS\,5 & 9.27 & 1.57 & I & 28. & 4,7
 & S & $HK_sL'$ & $\sim0.30$ & 500 & 04/12/03 \\
04295+2251 & L1536 IRS & 9.55 & 0.11 & FS & 0.6 & 4,7 & S & $K_sL'$ &
 0.39 & 900 & 04/12/03 \\
04361+2547 & TMR\,1 & 10.55 & 1.27 & I & 3.7 & 4,7 & B & $K_sL'$ &
 $\sim0.10$ & 140 & 04/12/03 \\
04365+2535 & TMC\,1A & 10.62 & 1.08 & I & 2.4 & 4,7 & S & $K_sL'$ &
 0.37 & 2800 & 04/12/03 \\
04385+2550 & Haro\,6-33 & 9.65 & 0.14 & FS & $>$0.4 & 4,7 & S &
 $HK_sL'$ & 0.67 & 2800 & 03/12/03 \\
04489+3042 & & 10.11 & 0.18 & FS & 0.3 & 4,7 & S & $K_sL'$ & 0.63 &
2200 & 04/12/03 \\
\hline
\multicolumn{12}{|c|}{Ophiuchus}\\
\hline
Oph\,29 & GSS\,30 & 8.32 & 1.20 & I & 21. & 5,6 & S & $HK_sL'$ & 0.33
& 5000 & 01/06/03 \\
Oph\,31 & LFAM\,1 & 13.59 & 1.08 & I & 0.13 & 3,6 & S & $K_sL'$ & 0.57
& 35 & 19/06/04, 11/04/04 \\
Oph\,46 & VSSG\,27 & 10.72 & 0.17 & FS & 0.41 & 5,6 & T & $HK_s$ &
0.35 & 1250 & 19/06/04 \\
Oph\,51 & & 9.59 & -0.09 & FS & 0.71 & 6,11 & B & $K_sL'$ & 0.17 & 750
& 02/06/03 \\
Oph\,108 & EL\,29 & 7.54 & 0.98 & I & 26.  & 5,6 & S & $HK_sL'$ &
$\sim0.65$ & 20000 & 01/06/03 \\
Oph\,121 & WL\,20 & 9.21 & 1.67 & I & 1.5 & 5,6,8 & S & $K_sL'$ & 0.62
& 90 & 27/03/05, 07/06/05 \\
Oph\,132 & IRS\,42 & 8.41 & 0.08 & FS & 5.6 & 5,6 & S & $K_sL'$ & 0.52
& 5000 & 02/06/03 \\
Oph\,134 & WL\,6 & 10.04 & 0.59 & I & 1.7 & 5,6 & S & $K_sL'$ & 0.42 &
4000 & 01/06/03 \\
Oph\,141 & IRS\,43 & 9.46 & 0.98 & I & 6.7 & 5,6 & B & $HK_sL'$ & 0.27
& 1500 & 01/06/03 \\
Oph\,143 & IRS\,44 & 9.65 & 1.57 & I & 8.7 & 5,6 & B & $K_sL'$ &
$\sim0.30$ & 1900 & 30/04/05, 09/04/05 \\
Oph\,147 & IRS\,47 & 8.95 & 0.17 & FS & 3.7 & 5,6 & B$^\star$ &
$HK_sL'$ & $\sim0.40$ & 2800 & 02/06/03 \\
Oph\,159 & IRS\,48 & 7.42 & 0.18 & FS & 7.4 & 5,6 & S & $HK_sL'$ &
0.29 & 2000 & 01/06/03 \\
Oph\,167 & IRS\,51 & 8.93 & -0.04 & FS & 1.1 & 5,6 & B & $HK_sL'$ &
0.41 & 1500 & 01/06/03 \\
Oph\,182 & IRS\,54 & 10.87 & 1.76 & I & 6.6 & 5,6 & B & $K_sL'$ & 0.19
& 700 & 02/06/03 \\
Oph\,200 & & 10.43 & 0.22 & FS & 1.3 & 6,11 & S & $K_sL'$ & 0.50 &
2500 & 12/04/05 \\
Oph\,204 & L1689 IRS\,5& 7.90 & -0.25 & FS & 2.4 & 3,6 & T & $HK_sL'$
& 0.45 & 2200 & 02/06/03 \\
\hline
\multicolumn{12}{|c|}{Serpens$^{\mathrm{e}}$}\\
\hline
Ser\,159 & & 8.58 & 0.11 & FS & (\dots) & 9,11 & S & $HK_sL'$ & 0.67 &
1500 & 01/06/03 \\
Ser\,307 & SVS\,2 & 8.95 & 0.06 & FS & (\dots) & 9,11 & S & $HK_sL'$ &
0.48 & 1200 & 01/06/03 \\
Ser\,312\,A & EC\,88 & 13.38 & 1.70 & I & 38. & 10,11 &
T$^\star$ & $K_sL'$ & 0.49 & 2900 & 30/04/05 \\
%
Ser\,314 & SVS\,20 & 7.05 & -0.03 & FS & (\dots) & 9,11 & T$^\star$ &
$HK_sL'$ & 0.66 & 6000 & 01/06/03 \\
Ser\,317\,A & EC\,92 & 10.50 & 0.50 & FS & 1.2 & 10,11 & S &
$K_sL'$ & 0.63 & 9500 & 22/05/05, 11/05/05 \\
%
Ser\,318 & EC\,94 & 11.65 & -0.05 & FS & 4.6 & 9,10,11 & S & $K_sL'$ &
0.63 & 800 & 22/05/05, 11/05/05 \\
Ser\,326 & EC\,103 & 11.84 & 0.62 & I & (\dots) & 9,11 & S & $L'$ &
0.58 & 2000 & 11/05/05 \\
Ser\,347 & EC\,129 & 9.92 & 0.06 & FS & (\dots) & 9,11 & S & $HK_sL'$
& 0.23 & 1200 & 01/06/03 \\
\hline
\multicolumn{12}{|c|}{L1641 (Orion\,A)}\\
\hline
IRAS\,29 & & 9.8 & 1.30 & I & 32.2 & 2 & B$^\star$ & $HK_sL'$ &
$\sim0.65$ & 2000 & 03/12/03 \\
IRAS\,50 & & 8.14 & -0.06 & FS & 99.6 & 1,2 & T$^{\star\star}$ &
$HK_sL'$ & $\sim0.50$ & 10000 & 03/12/03 \\
IRAS\,72 & & 9.49 & 0.15 & FS & 3.2 & 1,2 & B$^\star$ & $HK_sL'$ &
$\sim0.50$ & 1600 & 04/12/03 \\
IRAS\,79 & V883\,Ori & 4.98 & 0.23 & FS & 241. & 1,2 & S & $HK_s$ &
0.12 & 13000 & 04/12/03 \\
IRAS\,87 & & 10.7 & 0.56 & I & 10.2 & 2 & S & $K_sL'$ & 0.54 & 1800 &
04/12/03 \\
IRAS\,120$^{\mathrm{f}}$ & & 8.97 & -0.43 & II & 1.5 & 1,2 & S &
$HK_sL'$ & 0.74 & 2000 & 04/12/03 \\
IRAS\,171\,A & & 9.60 & -0.05 & FS & 2.4 & 1,2 & S & $HK_sL'$ & 0.66 &
500 & 03/12/03 \\
IRAS\,171\,B & & 10.1 & -0.05 & FS & 2.4 & 1,2 & B$^\star$ & $HK_sL'$
& 0.66 & 100 & 03/12/03 \\
IRAS\,187 & V1791\,Ori & 8.11 & 0.01 & FS & 7.5 & 1,2 & S & $HK_sL'$ &
0.70 & 8000 & 03/12/03 \\
IRAS\,191 & DL\,Ori & 9.38 & 0.64 & I & 8.1 & 1,2 & S & $HK_sL'$ &
0.74 & 4000 & 03/12/03 \\
IRAS\,224 & & 10.18 & 0.12 & FS & 1.5 & 1,2 & S & $K_sL'$ & 0.39 & 700
& 04/12/03 \\
IRAS\,237 & & 9.5 & -0.21 & FS & 7.3 & 2 & S & $K_sL'$ & 0.50 & 7000 &
04/12/03 \\
IRAS\,270 & & 9.8 & 0.47 & I & 5.4 & 2 & B$^\star$ & $HK_sL'$ &
$\sim0.50$ & 3000 & 03/12/03 \\
\hline
\end{tabular}
\begin{list}{}{}
\item[$^{\mathrm{a}}$] References: 1) Strom et al. (1989); 2) Chen \&
  Tokunaga (1994); 3) Greene et al. (1994); 4) Kenyon \& Hartmann
  (1995); 5) Barsony et al. (1997); 6) Bontemps et al. (2001); 7)
  Motte \& Andr\'e (2001); 8) Ressler \& Barsony (2001); 9) Kaas et
  al. (2004); 10) Pontoppidan et al. (2004); 11) 2MASS Point Source
  Catalog.
\item[$^{\mathrm{b}}$] Multiplicity status, counting only candidate
  companions within 1400\,AU of their primary (i.e., excluding likely
  non-physical systems and candidate background companions). Asterisks
  indicate companions newly discovered in this survey (see
  Table\,\ref{tab:newcomps}).
\item[$^{\mathrm{c}}$] When multiple observation dates are indicated,
  they are listed in the same order as the filters they correspond to.
\item[$^{\mathrm{d}}$] IRAS\,04287+1801 has been resolved into a
  multiple system in radio observations (Rodr\'{\i}guez et al. 1998;
  Lim \& Takakuwa 2006) but only one of these components is detected
  in near-infrared images.
\item[$^{\mathrm{e}}$] Bolometric luminosities are not available for
  all sources in Serpens due to source confusion in the far-infrared
  (Kaas et al. 2004; see Pontoppidan et al. 2004).
\item[$^{\mathrm{f}}$] The Class\,II source IRAS\,120 is listed here
  for completeness only and is not used in our statistical analyses of
  Class\,I and FS protostars.
\end{list}
\end{table*}

We defined our sample on the basis of IRAS and ISO mid-infrared (IR)
surveys of the clouds. Specifically, we used Kenyon \& Hartman (1995)
in Taurus, Bontemps et al. (2001) in Ophiuchus, Kaas et al. (2004) in
Serpens and Chen \& Tokunaga (1994) in L1641. Throughout this paper,
we use the source naming conventions of these authors for the last
three clouds (e.g., ISOCAM numbers in Ophiuchus and Serpens) and the
IRAS Point Source Catalog names in Taurus. The protostars in our
sample are classified as Class\,I or ``flat spectrum'' (FS) on the
basis of their spectral index in the near- to mid-IR region, a
well-documented proxy for the evolutionary status of YSOs (Greene et
al. 1994). Specifically, we used $\alpha^{2-12}$ indices from $K$-band
photometry and IRAS 12\,$\mu$m fluxes in Taurus and Orion. One source
in L1641 (IRAS\,171) has two possible counterparts, only separated by
6\arcsec, which have almost identical $HKL'$ colors and are likely in
a similar early evolutionary state (see Chen \& Tokunaga 1994); we
consider them as two separate targets in our analysis. In Ophiuchus
and Serpens, we calculated $\alpha^{2-14}$ indices from Bontemps et
al. (2001) and Kaas et al. (2004). Ser\,312 and Ser\,317 both have two
near-IR counterparts, which correspond to a single entry in our
survey because they are separated by less than 1400\,AU, the upper
limit for binary separations considered here. To obtain independent
mid-IR fluxes and spectral indices for their individual
components, we used the refined analysis of Pontoppidan et
al. (2004). Following Greene et al. (1994), we consider that sources
with $\alpha_{IR}>0.3$ are Class\,I sources, sources with
$-0.3\leq\alpha_{IR}\leq0.3$ are FS, and sources with
$\alpha_{IR}<-0.3$ are Class\,II sources. The index and source
classification for each target, as well as their estimated bolometric
luminosities are listed in Table\,\ref{tab:sample_obs}. In Serpens,
source confusion at the longest wavelengths prevents assigning
bolometric luminosities to most individual sources; these are however
available for a subset of sources, based on the analysis of
Pontoppidan et al. (2004). No individual source should have a
luminosity higher than about 50$L_\odot$, however. Because Serpens and
Orion are further away then Taurus and Ophiuchus, the median
bolometric luminosity is on order of 4--5$L_\odot$ instead of
1--2$L_\odot$ in the closer clouds. While this may induce a systematic
bias from one cloud to the other, we note that our sample consists of
objects whose mass is likely $\lesssim2\,M_\odot$, i.e., low-mass
protostars, except for IRAS\,50 and IRAS\,79.

We note that the use of only an IR spectral index to classify YSOs may
not be the most physical criterion. In particular, Class\,I protostars
are primarily defined by the fact that they are surrounded by a
substantial remnant envelope (Lada 1987; Andr\'e \& Montmerle
1994). It is possible that some of the targets in our sample do not
match this criteria, their spectral index being affected by source
confusion, variability or a particular geometry, for instance (see
discussion in Section\,\ref{sect:results}.1.1). Also, Class\,I YSOs
can be strongly variable at near-IR wavelengths, shading doubt on
their actual nature. In the absence of a more refined criteria
available {\it for all targets}, we use the spectral index to define
our sample of ``embedded protostars'' and use this terminology
throughout this paper. We note, however, that variability or
photometric errors on the order of 20\% on each measurement can lead
to uncertainties of $\sim0.15$ on the spectral index, so that objects
close to a transition value may be classified differently in other
studies. The case of IRAS\,120 is particularly striking in this
respect: discrepancies of up to 1\,mag in the $K$ band, equivalent to
changes of spectral index of up to 0.5, have been documented in the
past (Strom et al. 1989; Chen \& Tokunaga 1994; 2MASS Point Source
Catalog). Similarly, Ser\,317\,B (EC\,95) has been regularly referred
to as a FS source (e.g., Preibisch 2003), although the photometry of
Pontoppidan et al. (2004) and Haisch et al. (2006) suggest a Class\,II
classification. On the other hand, recent Spitzer photometry indicate
that this source is rather a Class\,I source (Winston et
al. 2007). Both variability and crowding in the rich SVS\,4 area can
be held responsible for such uncertainty on the nature of
EC\,95. Although this is arguable, we decided to use homogeneous
datasets for all sources of a given cloud, as far as possible. On the
basis of Chen \& Tokunaga (1994) in Orion and Pontoppidan et
al. (2004) in Serpens, we do not include IRAS\,120 and Ser\,317\,B in
our analysis; our results on these targets are indicated for
completeness only. Finally, we emphasize that our targets do not
represent the earliest stage of stellar evolution, in which the
envelope mass exceeds that of the central source. However, Class\,0
sources are usually not detected in the near-IR and are not part of
this study.

The objects in our sample are distributed throughout their parent
clouds, without particular clustering, except for the SVS\,4 group in
the Serpens cloud which contains Ser\,312\,A and B, Ser\,317\,A and B
and Ser\,318. Even in Orion, which produced the ONC, one of the
densest stellar clusters in the solar neighborhood, our sample rather
represents a distributed population throughout the L1641 cloud.

Due to our choice of instrument (see below) and extremely red colors
of our targets in the visible and near-IR, we were limited to
$K\approx11$ for direct acquisition of high-resolution images. Due to
time limitations, we observed all but a few known protostars brighter
than this limit. On the other hand, we also observed a handful of
fainter sources (1 in Ophiuchus, and 4 in Serpens) using a brighter
nearby target as guide star. Our sample of 45 targets is therefore not
exactly flux-limited. Rather, considering the lists of objects from
Kenyon \& Hartmann (1995), Bontemps et al. (2001), Kaas et al. (2004)
and Chen \& Tokunaga (1994), our survey covers roughly one third of
all known Class\,I and FS targets in each molecular cloud, focusing
primarily on the brightest objects.

\begin{figure*}
\caption{\label{fig:contours}$L'$-band contour plots of all
  subarcsecond candidate companions to Class\,I and FS sources
  discovered in this survey. The contours are spaced by
  0.75\,mag/arcsec$^2$ from $\approx$90\% of the the peak flux down to
  approximately the 5$\sigma$ level. All images are 3\arcsec\ on a
  side. In each panel, a scale bar represents 100\,AU at the distance
  of the target. In the highest $SR_{L'}$ images, particularly for
  Ser\,314 and IRAS\,29, PSF substructures could be confused with
  companions. These features are in fact adaptive optics artifacts
  which tend to appear at constant locations, allowing for their
  identification. Such strong artifacts are also seen along an
  horizontal line and both diagonals in the images of IRAS\,72 and
  IRAS\,270.}  \includegraphics[angle=-90,width=\textwidth]{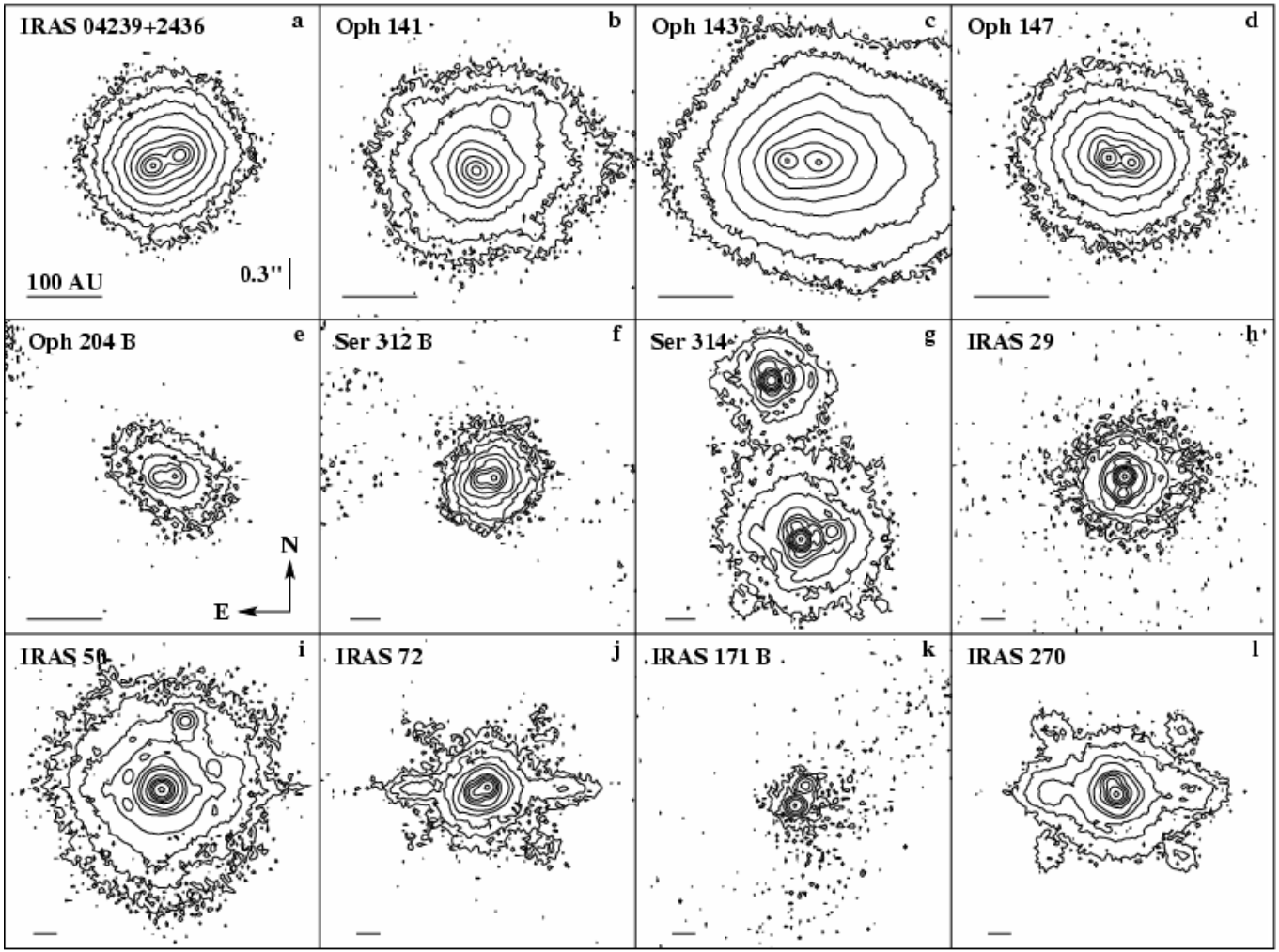}
\end{figure*}


\subsection{Observations and data reduction}

The observations were conducted with the Nasmyth Adaptive Optics
System (NAOS) installed on the Yepun 8.2\,m-Unit Telescope at ESO's
Very Large Telescope, in combination with the CONICA instrument
(Rousset et al. 2002; Lenzen et al. 2003). We used the 0\farcs0272
pixel scale, yielding a total field-of-view of
28\arcsec$\times$28\arcsec. The data were acquired in Visitor Mode in
June and December 2003 and in Service Mode between April to June 2004
and between March to June 2005. The median seeing was 0\farcs75, but
varied between 0\farcs4 and 1\farcs25 from one observation to
another. Sky conditions varied from photometric to transparent. A
detailed observing log is presented in Table\,\ref{tab:sample_obs}.

Because our targets are undetected in the optical, we used the
IR wavefront sensor of NAOS, with two dichroics: the so-called
N90C10 for $H$ (1.66\,$\mu$m) and $K_s$ (2.18\,$\mu$m) imaging and the
so-called JHK for $L'$ (3.80\,$\mu$m) imaging; their effective
transmission to the detector is on the order of 5--10\% at $H$ and
$K_s$ and 90\% at $L'$. All targets, except Oph\,46 and IRAS\,79, were
imaged with the $L'$ filter. All but Ser\,326 were further observed
with the $K_s$ filter and about 60\% of the targets with the $H$
filter. At $L'$ several tens to 200 short exposures were coadded to
reach a total integration time of 10--30\,sec without saturating the
background. At $H$ and $K_s$, a few (typically 5 to 30) longer
exposures were similarly coadded to reach a similar signal-to-noise
ratio (SNR). This sequence was repeated 4 to 10 times, each time after
randomly jittering the location of the stars by a few arcseconds on
the chip to better correct for bad pixels. The exact integration times
and number of coadd images and jitter positions depend on the object
brightness and the quality of the adaptive optics (AO)
correction. This is measured by the achieved Strehl ratio (SR), which
is the ratio of the peak pixel by the encircled energy divided by the
same ratio for a perfect AO correction, namely an Airy function. We
list in Table\,\ref{tab:sample_obs} the SR measured in the $L'$ image
(or $K_s$ image if the $L'$ filter was not used), as well as the peak
SNR calculated as the ratio of the value of the target's peak pixel
and the r.m.s. dispersion in the background.

The data reduction was performed with {\sl eclipse} (Devillard 2001)
and consisted of the usual steps. First, all images were flat-fielded
using flat fields taken as part of the ESO's Calibration Plan. A sky
was then created by clipped-median-combining all images of an object
obtained with the same filter and that sky was subtracted from all
images. A cosmetic correction was then applied for bad pixels and
cosmic rays. Finally, the images were shift-and-added to produce the
final images used in this survey. Contour plots for all sub-arcsecond
companions detected in this survey are presented in
Figure\,\ref{fig:contours}.

\begin{figure}
\caption{\label{fig:detect}Companions to Class\,I and FS protostars
  surveyed at high angular resolution in our survey and in the
  literature (Tables\,\ref{tab:newcomps} and \ref{tab:extra_sample}
  and D04). Filled squares represent candidate companions to
  low-luminosity Class\,I and FS sources, triangles companions to
  high-luminosity protostars in Orion ($L_{bol} \gtrsim100 L_\odot$),
  empty squares likely non-physical systems and crosses probable
  background companions. For sources surrounded by a diamond, the flux
  ratio was measured at $K_s$ in the absence of an $L'$ image. The
  solid lines are 3$\sigma$ detection limits up to 2\farcs7 (i.e., the
  background limit) for 4 single stars observed in our survey with a
  very similar $SNR_{L'}$ (2000--2200) but with a varying quality of
  AO correction at L-band; from top to bottom at 0\farcs15, the stars
  are Oph\,159 ($SR_{L'}=29\%$, close to worst in our dataset),
  Oph\,204 ($SR_{L'}=45\%$), Ser\,159 ($SR_{L'}=67\%$) and IRAS\,120
  ($SR_{L'}=74\%$, best in our dataset).}
  \includegraphics[width=\columnwidth]{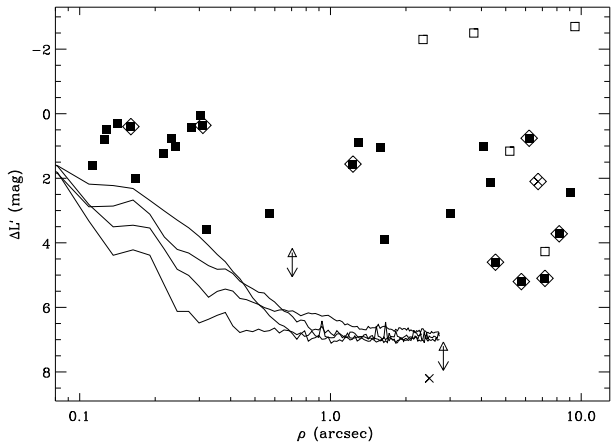}
\end{figure}

From objects which appear point-like in our images and that were
observed with a good AO correction (SR$\gtrsim$50\%, the median SR in
our observations), we estimate that the FWHM of the point spread
function (PSF) is about 0\farcs060, 0\farcs070 and 0\farcs110 at $H$,
$K_s$ and $L'$, respectively. Because our survey was primarily
conducted at $L'$ but with $K_s$ images available for almost all
targets, we conservatively consider that we could unambiguously detect
equal-flux companions down to 0\farcs1, which we consider the lower
limit for this survey, except for a few cases which appear extended in
our images; these are discussed in Section\,\ref{sect:specials}. Our
ability to detect companions depends on separation, flux ratio and
quality of the AO correction. To illustrate this, we calculated our
detection limit for companions as the 3$\sigma$ level in concentric
annuli around 4 single stars. These stars were selected because they
were observed with a very similar $SNR_{L'}$ (2000--2200, roughly the
median value in our observations), but with markedly different levels
of quality in the AO correction ($SR_{L'}$ ranging from 29 to
74\%). As can be seen in Figure\,\ref{fig:detect}, the detection limit
drops gradually with a slope that depends mostly on $SR_{L'}$. At a
separation of 0\farcs1, companions can be detected down to $\Delta
L'\approx2$\,mag independently of the AO correction quality; at
0\farcs5, the detection limit drops to at least $\Delta
L'\approx5$\,mag. Outside 1--2\arcsec, the detection limit is
dominated by background noise and remains constant. For stars observed
with $SNR\approx2000$, we could detect distant companions down to
$\Delta L' < 7$\,mag. Note that all the companions we detected are
above the detection limit for our worse detection limit, as can be
seen in Figure\,\ref{fig:detect}, implying that the risk of a bias
induced by the varying quality of our AO images is very limited.

The only object for which the AO correction was too poor to include
this dataset in our survey is IRAS\,04361+2547 ($SR_{L'}\sim10\%$). We
failed to resolve the 0\farcs31 binary identified by Terebey et
al. (1998, 2001) on the basis of HST/NICMOS images. While it is not
possible to determine from our dataset whether this source is
intrinsically extended, we will take advantage of the HST images for
this target, as well as for a few other objects (see
Sect.\,\ref{sect:add_data}).

Relative astrometry and photometry for all companions detected in our
survey was performed with the DAOPHOT package within IRAF. We only
considered companions within 1400\,AU in projected separation of our
targets, in line with our analysis of D04 and with previous
multiplicity surveys among YSOs (e.g., Leinert et al. 1993; Simon et
al. 1995). When the companions were wider than about 1\arcsec,
relative photometry was performed using small enough apertures. For
tighter systems, PSF-fitting was performed, using another star located
in the field-of-view when available or another single target observed
with a similar AO correction. We estimate that our typical astrometric
uncertainty is on order 0.5\% for the separation and 0\fdg5 for the
orientation. Uncertainties on the relative photometry are on order
0.05--0.1\,mag, depending on the system's separation, flux ratio and
image quality. We assume that the brightest component at $L'$ is the
primary of the system, except for the wide systems Oph\,31, Oph\,121,
Ser\,312 and Ser\,317, for which previous observations allow
unambiguous identification of the components. In these cases, the
relative astrometry and photometry is presented using the Class\,I or
FS protostar, which is fainter than its companions in some images, as
the primary. The flux ratio and separation of all companions detected
in this survey are presented in Table\,\ref{tab:newcomps} and in
Figure\,\ref{fig:detect}. Most of the wide systems were known from
previous surveys (Haisch et al. 2002, 2004; D04) as well as from the
2MASS database. The tight companion to Oph\,204\,B was first found by
Ratzka et al. (2005). We note that the relative fluxes measured for
some systems in Taurus and Ophiuchus (IRAS\,04113+2758,
IRAS\,04263+2426, Oph\,31, Oph\,46, Oph\,121) differ from the
measurements of D04 by up to a magnitude, indicating that these
sources are variable. Such near-IR variability is typical of Class\,I
YSOs (e.g., Park \& Kenyon 2002).

\begin{table}
\caption{\label{tab:newcomps}Companions detected during this survey
  within 1400\,AU of their primary.}
\centering
\begin{tabular}{|l|cc|ccc|l|}
\hline 
Target & $\rho$ & P.A. & $\Delta H$ & $\Delta K_s$ & $\Delta L'$ &
Note$^{\mathrm{a}}$ \\
 & (\arcsec) & (\degr) & (mag) & (mag) & (mag) & \\ 
\hline
04113+2758 & 4.06 & 153.8 & -0.43 & 0.24 & $\sim$1.0$^{\mathrm{b}}$ &
 \\
04239+2436 & 0.279 & 326.9 & $>$1.7$^{\mathrm{c}}$ & 1.07 & 0.42 & 1
\\
04263+2426 & 1.29 & 173.7 & -3.2 & -1.57 & $\sim$0.9$^{\mathrm{b}}$ &
 \\
\hline
Oph\,31 & 9.44 & 105.7 & (\dots) & -5.2 & -2.7 & 2 \\
Oph\,46 & 1.23 & 66.4 & 1.15 & 1.56 & (\dots) & \\
 & 5.77 & 96.5 & 4.9 & 5.2 & (\dots) & \\
Oph\,51 & 9.08 & 152.5 & (\dots) & 0.37 & 2.45 & \\
Oph\,121 & 2.34 & 352.8 & (\dots) & -3.1 & -2.3 & 2 \\
 & 3.73 & 51.5 & (\dots) & -3.3 & -2.5 & 2 \\
Oph\,141 & 0.57 & 334. & $>$3.0$^{\mathrm{c}}$ & $>$4.0$^{\mathrm{c}}$
 & 3.1 & \\
 & 7.17 & 323.0 & 1.30 & 2.90 & 4.27 & 2 \\
Oph\,143 & 0.302 & 87.0 & (\dots) & $<$-1.4$^{\mathrm{c}}$ & 0.05 & \\
Oph\,147 & 0.232 & 258.2 & 0.25 & 0.51 & 0.76 & 1 \\
Oph\,167 & 1.64 & 10.1 & 3.35 & 3.56 & 3.90 & \\
Oph\,204\,AB & 3.02 & 240.4 & 1.83 & 3.09 & 3.08 & \\
Oph\,204\,B & 0.142 & 93.3 & -0.18 & -0.18 & 0.29 & \\
\hline
Ser\,312\,AB & 4.36 & 11.7 & (\dots) & -0.55 & 2.14 & \\
Ser\,312\,B & 0.128 & 92. & (\dots) & 1.60 & 0.50 & 1 \\
Ser\,314 & 0.321 & 284.3 & $>$4.0$^{\mathrm{c}}$ & 4.0 & 3.60 & 1
\\
 & 1.58 & 10.2 & 1.35 & 1.28 & 1.04 & \\
Ser\,317\,AB & 5.19 & 173.6 & (\dots) & 0.24 & 1.16 & 2 \\
Ser\,317\,A & 2.48 & 109.2 & (\dots) & 7.3 & 8.2 & 1,3 \\
Ser\,317\,B & 0.152 & 227.2 & (\dots) & -1.2 & 0.05 & 1,4 \\
\hline
IRAS\,29 & 0.167 & 173.8 & $>$2.5$^{\mathrm{c}}$ &
$>$2.5$^{\mathrm{c}}$ & 2.0 & 1 \\
IRAS\,50 & 0.704 & 341.1 & $>$6.5$^{\mathrm{c}}$ &
$>$7.0$^{\mathrm{c}}$ & $>$4.3$^{\mathrm{b}}$ & 1 \\
 & 2.82 & 144.3 & $>$7.5$^{\mathrm{c}}$ & $>$7.5$^{\mathrm{c}}$ &
$>$7.2$^{\mathrm{b}}$ & 1 \\
IRAS\,72 & 0.125 & 115.1 & -0.4 & 0.1 & 0.8 & 1 \\
IRAS\,171\,B & 0.215 & 333.5 & 0.94 & 1.14 & 1.22 & 1 \\
IRAS\,270 & 0.112 & 20.8 & 0.7 & 1.1 & 1.6 & 1 \\
\hline
\end{tabular}
\begin{list}{}{}
\item[$^{\mathrm{a}}$] 1: newly identified companion; 2: candidate
  non-physical system; 3: candidate background companion; 4: not part
  of the survey (given for completeness only).
\item[$^{\mathrm{b}}$] At least one of the two components is saturated
  in the image so that relative photometry cannot be accurately
  extracted.
\item[$^{\mathrm{c}}$] The companion is not detected and only an upper
  limit to its flux can be determined, based on its known location.
\end{list}
\end{table}


\subsection{Additional datasets}
\label{sect:add_data}

To complement our survey, we searched the literature for uniform high
angular resolution surveys of protostars in the clouds studied
here. We found 14 targets in Taurus and Ophiuchus observed with
HST/NICMOS and/or ground-based near-IR AO observations on a 4m-class
telescope, representing a total of 5 companions\footnote{We have
estimated in D04 that the faintest companion to Oph\,33 is a likely
background source and we exclude from our analysis.} within
1400\,AU. These observations are summarized in
Table\,\ref{tab:extra_sample}. Although other high-resolution images
on specific targets have been published (e.g., IRAS\,04325+2402,
Hartmann et al. 1999), we do not include them in our analysis to avoid
a likely bias towards binary/multiple systems. Two of the targets
studied by Padgett et al. (1999), DG\,Tau\,B and CoKu\,Tau\,1, are
candidate edge-on disks and it is unclear whether the central sources
are intrinsically Class\,I or Class\,II sources. We nonetheless
maintain these sources in our sample on the basis of their observed
IR spectral index, as we do for Oph\,31 (see
Section\,\ref{sect:specials}). We note that the survey of Allen et
al. (2002) was conducted with a 0\farcs2 pixel scale, resulting in a
poor capacity to detect companions closer than 0\farcs5. It is
therefore possible that a few very tight companion were missed in that
survey. Furthermore, the astrometry listed in Allen et al. (2002)
appears inaccurate: the separations of the wide companions to Oph\,33
and Oph\,141 are 15--20\% too large and 18\% too small,
respectively. We do not know the reason for this discrepancy. For
Oph\,33 we adopt the astrometry presented in D04 and for Oph\,141 that
obtained during this survey, which is consistent with D04.

\begin{table*}
\caption{\label{tab:extra_sample}Additional protostars observed at
high resolution from the literature.}  \centering
\begin{tabular}{|ll|ccccc|cc|ccl|l|}
\hline
 Target & Alt. Name & $K$ & $\alpha_{\mathrm IR}$ & Class &
 $L_{\mathrm bol}$ & Ref.$^{\mathrm{a}}$ & $\rho$ & P.A. & $\Delta H$
 & $\Delta K_s$ & Ref.$^{\mathrm{a}}$ & Note$^{\mathrm{b}}$ \\
 & & & & & $(L_\odot)$ & & (\arcsec) & (\degr) & (mag) & (mag) & & \\
\hline
\multicolumn{13}{|c|}{Taurus}\\
\hline
04016+2610 & & 9.33 & 1.06 & I & 3.7 & 2,7 & & & & & 5 & \\
04169+2702 & & 11.22 & 0.89 & I & 0.8 & 2,7 & & & & & 8 & \\
04248+2612 & & 10.65 & 0.48$^{\mathrm{c}}$ & I & 0.4 & 2,7 &
0.160 & 266. & 0.20 & 0.40 & 5 & \\
 & & & & & & & 4.55 & 15.1 & (\dots) & 4.60 & 10 & \\
04302+2247 & & 11.52 & 0.15$^{\mathrm{c}}$ & FS & 0.3 & 2,7 & &
& & & 5 & \\
04361+2547$^{\mathrm{d}}$ & TMR\,1 & 10.55 & 1.27 & I & 3.7 &
2,7 & 0.31 & 19. & 0.88 & 0.36 & 4,8 & \\
04381+2540 & TMC\,1 & 12.00 & 1.20 & I & 1.31 & 2,6 & & & & & 8
& \\
DG\,Tau\,B & & 11.52 & 2.05$^{\mathrm{e}}$ & I & & 2,12 & & & & & 5 &
1 \\
CoKu\,Tau\,1 & & 10.85 & 0.82$^{\mathrm{e}}$ & I & 0.065 & 1,2,12 &
0.240 & 111. & 1.32 & 1.03 & 5 & 1 \\
\hline
\multicolumn{13}{|c|}{Ophiuchus}\\
\hline
Oph\,33 & GY\,11 & 14.15 & 0.31 & I & 0.011 & 3,6 & 6.20 & 86.1 & 1.2
& 0.76 & 9,10 & \\
 & & & & & & & 6.73 & 40.9 & 3.1 & 2.10 & 9,10 & 2 \\
Oph\,37 & LFAM\,3 & 9.94 & -0.06 & FS & 0.58 & 3,6 & & & & & 9,10 & \\
Oph\,85 & CRBR\,51 & 14.00 & 0.03 & FS & 0.035 & 3,6 & & & & & 9 & \\
Oph\,103 & WL\,17 & 10.28 & 0.42 & I & 0.88 & 3,6 & & & & & 9 &
\\
Oph\,137 & CRBR\,85 & 13.21 & 1.48 & I & 0.36 & 3,6 & & & & & 9 &
\\
Oph\,145 & IRS\,46 & 11.46 & 0.94 & I & 0.62 & 3,6 & & & & & 9 &
\\
\hline
\end{tabular}
\begin{list}{}{}
\item[$^{\mathrm{a}}$] References: 1) Strom \& Strom (1994); 2) Kenyon
  \& Hartmann (1995); 3) Barsony et al. (1997); 4) Terebey et
  al. (1998); 5) Padgett et al. (1999); 6) Bontemps et al. (2001); 7)
  Motte \& Andr\'e (2001); 8) Terebey et al. (2001); 9) Allen et
  al. (2002); 10) D04; 11) Ratzka et al. (2005); 12) Luhman et
  al. (2006).
\item[$^{\mathrm{b}}$] 1: edge-on disk source, for which the
  classification based on the $\alpha_{IR}$ index is doubtful; 2:
  candidate background companion.
\item[$^{\mathrm{c}}$] The spectral indices for IRAS\,04248+2612 and
  IRAS\,04302+2247 are calculated from their $K$ band and IRAS
  25\,$\mu$m photometry as there are only upper limits on their IRAS
  12\,$\mu$m fluxes.
\item[$^{\mathrm{d}}$] We observed IRAS\,04361+2547 in our survey but
  did not detect the previously known companion that is listed here.
\item[$^{\mathrm{e}}$] Due to confusion with bright nearby sources,
  there is no spatially resolved fluxes for DG\,Tau\,B and
  CoKu\,Tau\,1. The spectral indices listed here are values of
  $\alpha^{2-8}$ based on recent {\it Spitzer}/IRAC photometry.
\end{list}
\end{table*}

Combined with our VLT AO observations, these additional observations
allow us to construct a sample of 29 Class\,I and 29 FS objects
observed at high-angular resolution in the near-IR. This sample,
which represents 50--60\% of all known such sources in Taurus and
Ophiuchus, is the basis for the statistical analysis that we conduct
in the following.


\section{Results}
\label{sect:results}


\subsection{Notes on individual objects}
\label{sect:specials}

\subsubsection{Extended objects}

A few objects in our survey appeared extended in our AO images at
$K_s$ and $L'$, resulting in a much poorer ability to detect nearby
companions. While this is a consequence of a poor AO correction in the
case of IRAS\,04361+2547, the spatial extension of IRAS\,04287+1801,
Oph\,31, Oph\,121 and Oph\,143 is real, as shown by images of point
sources found in the field-of-view. These objects are discussed in
more details below.

\begin{figure}
\caption{\label{fig:extended}$L'$-band contour plot of
  IRAS\,04287+1801 and Oph\,31. The contours are spaced by
  0.75\,mag/arcsec$^2$ from $\approx$90\% of the the peak flux down to
  approximately the 5$\sigma$ level. Each image is 4\arcsec\ on a side.}
  \includegraphics[angle=-90,width=\columnwidth]{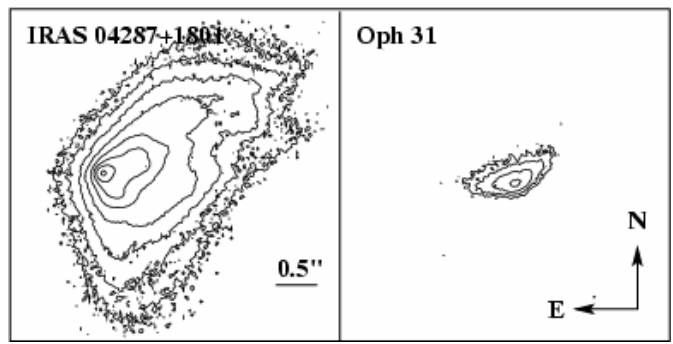}
\end{figure}

IRAS\,04287+1801 presents a central point-like source surrounded by a
vast nebulosity (Fig\,\ref{fig:extended}), similar to the case of
Oph\,143 (Fig\,\ref{fig:contours}c) and probably due to scattering off
a circumstellar envelope or the base of a wide outflow. Our ability to
detect tight companions is strongly reduced because of the enhanced
background and the known radio companion located 0\farcs35 South of
the primary\footnote{The third companion recently detected by Lim \&
Takakuwa (2006) at 7mm is so close from the primary (0\farcs09
projected separation) that we probably would not have detected it even
in the absence of the extended nebulosity.} is undetected in our
images (Looney et al. 1997; Rodr\'{\i}guez et al. 1998). Because it
has never been detected in the near-IR, this companion is not
included in our study, but this illustrates that some companions may
have been missed in our survey.

The apparent shape of Oph\,31 (LFAM\,1) appears different from the
vast nebulosities surrounding other Class\,I protostars. Most
noticeably, no central point source is found, even at $L'$
(Fig\,\ref{fig:extended}). Furthermore, the overall extent of Oph\,31
is much smaller. While this is in part due to the lower SNR of these
observations, the contours appear to drop sharply, suggesting a small
intrinsic size for this source\footnote{The HST/NICMOS images of Allen
et al. (2002) also show that Oph\,31 is a small, though clearly
resolved, source.}. Indeed, the elongated, slightly triangular shape of
Oph\,31 is reminiscent of images of T\,Tauri stars which possess an
edge-on opaque circumstellar disk, such as HH\,30 (Burrows et
al. 1996) or HK\,Tau\,B (Stapelfeldt et al. 1998), albeit with a
contrast between the main nebula and the counternebula that is too
high for us to detect the latter. Our $K_s$-band image reveals a
similar shape, though with a poorer SNR. If this is confirmed with
future, deep high-resolution imaging of the system, the possibility
that this object is an envelope-free Class\,II source whose rising SED
is the consequence of the presence of an edge-on disk will have to be
revisited. We note however that several of the best studied Class\,II
edge-on sources (HH\,30, HK\,Tau\,B, HV\,Tau\,C) have observed
spectral indices that are typical of normal Class\,II sources,
i.e. $\alpha^{2-12}\lesssim -0.5$, as a consequence of approximately
gray scattering (Stapelfeldt \& Moneti 1999; McCabe et al. 2006). Only
a very specific configuration, in which the edge-on disks becomes
optically thin between 2 and 10\,$\mu$m, may result in a
misclassification of a Class\,II source in a Class\,I source. This is
for instance the case of the newly discovered high-inclination disk
surrounding IRAS\,04158+2805, which has $\alpha^{2-12}= 0.37$
(M\'enard et al. 2007; Glauser et al. 2007). We do not have evidence
that this particular optical depth effect occurs for Oph\,31, and we
keep it in our sample for now on the basis of its very red spectral
index.

\begin{figure}
\centering
\caption{\label{fig:wl20s} $L'$-band image of the Oph\,121 triple
system. Both northern components, Class\,II sources, are point-like
and show the high quality of the AO correction, with marked Airy
rings. The blown-out panel of the Class\,I southern component reveals
a complex structure. Both panel use squared root stretches though with
different cuts to improve the visibility of features. The large image
and the inset are 8\arcsec\ and 2\arcsec\ on a side, respectively.}
\includegraphics[width=0.75\columnwidth]{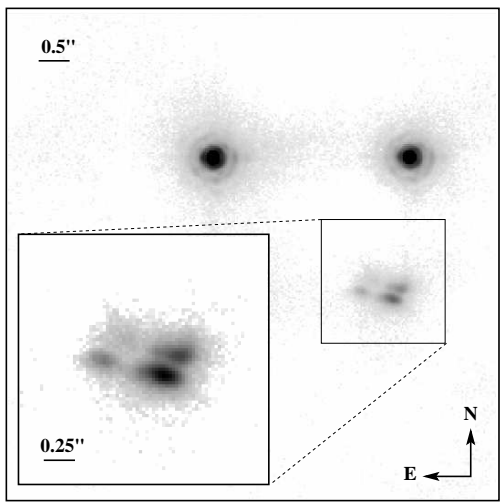}
\end{figure}

Oph\,121 (WL\,20\,S), the southwestern component of a triple system,
is also extended in our high-resolution images, as shown in
Figure\,\ref{fig:wl20s}. This objects shows four elongated, roughly
parallel nebula, most prominent at $L'$. The eastern, fainter nebulae,
are significantly redder than the western nebulae
($K_s-L'\approx2.6$--2.8 instead of $K_s-L'\approx1.9$--2.1). The two
nearby T\,Tauri stars, observed simultaneously, show that the AO
correction was good and that the spatial extension of Oph\,121 is
intrinsic. This object was studied in detail by Ressler \& Barsony
(2001), as an ``IR companion'' to 2 T\,Tauri stars. From
mid-IR images with 0\farcs3 spatial resolution, they found the
source to be extended on a scale of about 40\,AU, roughly the size of
the extended nebulosity in our images. The peak of the mid-IR
emission is not exactly coincident with any of the nebulae we found,
suggesting that we see only scattered light from a more embedded
source that can be seen directly in the mid-IR, a regime in
which dust absorbs much less. We defer a more detailed discussion of
the nature of this source to a future dedicated paper but note that
this could also be a Class\,II object that mimics a Class\,I source.

\subsubsection{The tight companions to Oph\,141 and Oph\,143}

Oph\,141 (IRS\,43) was resolved as a 0\farcs5 binary in the
mid-IR by Haisch et al. (2002), whose near-IR images did
not have enough spatial resolution to detect it. Furthermore, Terebey
et al. (2001) and Allen et al. (2002) failed to detect this companion
with HST/NICMOS. We clearly detect the companion at $L'$
(Figure\,\ref{fig:contours}b), with a relative astrometry consistent
with that given by Haisch et al. (2002), but fail to detect it at $H$
and $K_s$, as earlier HST surveys. Because of its point-source
appearance at $L'$, we nonetheless consider this companion as real.

Both Terebey et al. (2001) and Ratzka et al. (2005) resolved Oph\,143
(IRS\,44) as a binary system with a separation $\sim$0\farcs26 and a
position angle similar to that found in our $L'$ image
(Figure\,\ref{fig:contours}c and Table\,\ref{tab:newcomps}). We failed
to detect this companion at $H$ and $K_s$ over the bright extended
nebulosity surrounding the protostar but both components appear point
source-like in our $L'$ image, i.e., with a 0\farcs1 resolution. On
the other hand, Allen et al. (2002) found two spatially extended
structures separated by about 0\farcs5, still along the same position
angle (260\degr). Our $K_s$ image shows a bright, extended nebulosity
up to 0\farcs7 from the eastern component of the system, which may be
what Allen et al. (2002) detected; again, their images suffered from a
poor spatial sampling. We believe that the slight inconsistencies in
astrometry are the consequence of the fact that the brightest $L'$
component is embedded at $K_s$ and therefore it is difficult to
pinpoint its location in previous observations. We therefore consider
that this is the same companion, which is indeed the primary of the
system.

\subsubsection{The wide companions to Oph\,31, Oph\,121, and Oph\,141}

The wide companion to Oph\,31 (LFAM\,1) is Oph\,34 (GY\,12), a
Class\,III object (Bontemps et al. 2001), while the two companions of
Oph\,121 (WL\,20) are Class\,II objects (Ressler \& Barsony
2001). Similarly, the wide companion to Oph\,141 (IRS\,43) is GY\,263,
an apparent Class\,II source (Haisch et al. 2002). It is unlikely,
though not impossible, that an embedded protostar and a more evolved
pre-Main Sequence star form a physical system. This suggests that at
least some of these systems may result from a chance alignment of
two/three cloud members in a crowded area of the Ophiuchus cloud.

Although we have estimated in D04 that the probability that these
systems are physically bound is higher than 97\%, it must be
emphasized that such statistical tools are merely indicative when
considering individual systems. Whenever possible, physical arguments
should be given priority in confirming the physical nature of apparent
systems. On the basis of their evolutionary status, we conservatively
consider all these companions as non-physical in this study, but
remind the reader that it is possible that some systems are indeed
physically bound.

\subsubsection{The faint companions to Oph\,51 and Oph\,182}

In D04, we identified 2 faint companions to Oph\,51, located 6\farcs1
and 6\farcs3, respectively, that we did not detect in this
survey. Similarly, a faint 7\farcs2 companion to Oph\,182 (IRS\,54)
remained undetected in our higher-resolution images. In both cases,
this is due to their faintness, since all these companions have
$\Delta K>5$\,mag, and to the poor AO correction we obtained at
$K_s$. At $L'$, we can place an upper limit to the brightness of the
companions at $\Delta L'\gtrsim 6$\,mag. In the case of Oph\,51, both
of these companions have probabilities of being physically bound to
the protostar ($\sim85\%$) below the 2$\sigma$ level, and we consider
them as likely background objects. On the other hand, we estimated in
D04 the probability that the companion to Oph\,182 is physically bound
to be 97\%. In the absence of more stringent physical arguments, we
consider it as a candidate companion and include it in our analysis.

\subsubsection{Ser\,312, Ser\,314 and Ser\,317}

Ser\,312 (EC\,88--89) and Ser\,317 (EC\,92--95) are both located in
the rich SVS\,4 complex. Our images reveal 10 stars in a 30\arcsec\
square area. Most of these are known members of the Serpens clouds,
with a mixture of Class\,I, FS and Class\,II sources (Pontoppidan et
al. 2004). In such a crowded area, statistical arguments suggest that
wide pairs may not be physically bound. 

While Ser\,317\,A (EC\,92) is undoubtedly a Class\,I source, the
status of Ser\,317\,B (EC\,95) is more uncertain. This object is
frequently referred to as a Class\,I source in the literature (e.g.,
Preibisch 2003), although its spectral index is that of a Class\,II
source (Pontoppidan et al. 2004). Since we have decided to use
homogeneous estimates of the spectral index, we use the latter
classification in our study. We therefore believe that this is an
example of a non-physical system and we do not include it in our
survey. As a consequence, the newly identified tight binary system
Ser\,317\,B is not part of our studied sample either; its observed
properties are nonetheless listed in Table\,\ref{tab:newcomps} for
completeness. Ser\,317\,A itself possesses a 2\farcs5 companion which
is extremely faint and much bluer than its Class\,I primary. As
discussed in Section\,\ref{sect:props}, we believe that this is a
background star observed through the moderate extinction of the cloud
at that location ($A_V\sim10$\,mag; Schnee et al. 2005). Ser\,317 is
therefore counted as a single star in our survey. On the other hand,
despite the crowdedness of this region, we consider the wide
Ser\,312\,A--B pair as physical as both objects have a protostar-like
IR spectral index (Pontoppidan et al. 2004). Ser\,312\,B (EC\,89)
itself possesses a tight companion so that this system is counted as
triple in our survey.

Ser\,314 (SVS\,20) is a well-known wide binary system (Eiroa et
al. 1987). Our observations reveal that the primary of this system,
SVS\,20\,S, also possesses an additional tight companion which was not
known prior to this survey. This is therefore a triple system.


\subsection{Multiplicity rate}
\label{sect:rate}

\subsubsection{Defining Complete and Extended Surveys}

Within 1400\,AU of the 58 Class\,I and FS targets surveyed in our
sample, we have detected a total of 35 candidate companions, 12 of
which are new discoveries. Including the faint companion to Oph\,182
from D04 and excluding the probable background companion to Ser\,317
and the likely non-physical systems in Ophiuchus and Serpens (see
Sect.\,\ref{sect:specials}), we therefore have a total of 28 candidate
companions, i.e., a raw multiplicity rate of 48$\pm$7\% (uncertainties
are calculated assuming binomial statistics). Since our sample was
mostly defined on the basis of a flux limit, we may have biased our
results towards finding too high a multiplicity rate. If a target
slightly fainter than our $K=11$ limit has a tight and nearly equal
flux companion, the combined system may be bright enough to enter our
sample. In our sample, 46 targets have an unresolved photometry of
$K_{syst}<11$, representing 25 candidate companions. Only 3 binary
systems (IRAS\,04248+2612, IRAS\,04361+2547 and CoKu Tau\,1) have a
primary that is fainter than $K=11$, so that the multiplicity rate of
the subsample restricted to primaries brighter than this limit is
51$\pm$8\%, indistinguishable from the rate quoted above. The flux
bias is therefore negligible in our survey.

Because our targets lie at different distances from the Sun and
represent a mixed bag of objects, with bolometric luminosities ranging
from $\approx0.01\,L_\odot$ to 241\,$L_\odot$, it is difficult to
compare this raw multiplicity rate to other surveys. We must therefore
define a sample that is uniform in order to perform such
comparisons. First of all, we discard the 2 most luminous targets in
L1641 (IRAS\,50 and IRAS\,79, $L\gtrsim100L_\odot$), so that the most
luminous targets in all clouds have $L_{bol}\lesssim40L_\odot$, or
$M_\star \lesssim 2 M_\odot$. We then define a Complete Survey by
considering a range of projected separation of 45--1400\,AU. The lower
limit corresponds to 0\farcs1, our survey completeness limit, at the
distance of L1641, the cloud furthest away from the Sun. This aims at
circumventing the issue of the different distances to our targets. In
this Complete Survey, we found 18 companions to 56 targets (see
Table\,\ref{tab:stats}), i.e., a multiplicity rate of
32$\pm$6\%. Taking advantage of the highest resolution achieved in our
survey, we also define an Extended Survey that focuses on the Taurus
and Ophiuchus clouds and spans the separation range 14--1400\,AU. The
multiplicity rate in this Extended Survey is 47$\pm$8\% (18 companions
to 38 targets).

Although the number of targets in each cloud is limited, we also
consider the possibility of cloud-to-cloud differences in the
multiplicity rates. In our Complete Survey, the Ophiuchus, Serpens and
L1641 samples present the same multiplicity rate (average:
38$\pm$8\%), which is however 1.5$\sigma$ higher than that found in
Taurus (19$\pm$10\%), an effect reminiscent of the finding by Haisch
et al. (2004). Fischer's $t$ test indicates that the confidence level
associated to this difference is on the order of only 84\%. A deficit
of companions in the Taurus cloud is surprising since our larger-scale
survey in D04 showed that Taurus and Ophiuchus had indistinguishable
multiplicity rates in the similar separation range
110-1400\,AU. Furthermore, the difference between the Taurus and
Ophiuchus multiplicity rates in our Extended Survey is negligible
($\lesssim0.4\sigma$). While there seems to be a higher proportion of
companions in the 14--45\,AU separation range in Taurus than in
Ophiuchus, this is not statistically significant given the small size
of our samples (confidence level $\sim$50\% based on Student's $t$
test). We therefore conclude that there is no significant difference
in the multiplicity rates of Class\,I protostars across all four
molecular clouds studied here.

\begin{table}
\caption{\label{tab:stats}Number of companions in our Complete and
Extended Surveys.}
\centering
\begin{tabular}{|c|cccc|}
\hline
 Cloud & Taurus & Ophiuchus & Serpens & Orion \\
\hline
$N_{targets}$ & 16 & 22 & 8 & 10 \\
\hline
$N_{comp}$ (45--1400\,AU) & 3 & 8 & 3 & 4 \\
\hline
$N_{comp}$ (14--1400\,AU) & 7 & 11 & (\dots) & (\dots) \\
\hline
\end{tabular}
\end{table}

\subsubsection{Triple and higher order multiple systems}

Overall, we found 6 triple systems, including the high-luminosity
protostar IRAS\,50 in Orion, and no higher order system in our
survey. Within the Complete Survey, there are 40 single objects, 14
binaries and 2 triple systems. In other words, at least about 13\% of
all wide binaries are in fact higher order multiples. Since our survey
probed a limited range of separations, this is to be considered a
conservative lower limit to this ratio. For instance, Covey et
al. (2006) found a large difference in radial velocity between the
secondary of IRAS\,04263+2426 and the surrounding gas cloud and
suggested that the system may contain a third component. This
component would be in such a tight orbit that we could not have
resolved it here.

All triple systems in our surveys appear hierarchical: the ratio of
their projected separations is always larger than 4, reaching more
than 20 for IRAS\,04248+2612, Oph\,204 and Ser\,312. The long term
stability of triple systems depends on the ratio of semi-major axes,
on the mass ratios and on the eccentricities of the inner and outer
orbits (e.g., Eggleton \& Kiseleva 1995). None of these quantities is
directly available to us, but it is likely that most, if not all, of
the multiple systems we discovered are stable on the long term. On the
other hand, the fact that the Oph\,121 triple systems is apparently
non-hierarchical reinforces the suspicion that the companions to this
protostar are not physically bound to it (see
Section\,\ref{sect:results}.1.3).


\subsection{Colors of detected companions}
\label{sect:props}

\begin{figure}
\caption{\label{fig:ccd}Color-color diagram and color-magnitude
  diagram for protostars in our sample (crosses) and companions
  detected in this survey (filled squares: physical companions; empty
  squares: non-physical and background companions).}
  \includegraphics[width=\columnwidth]{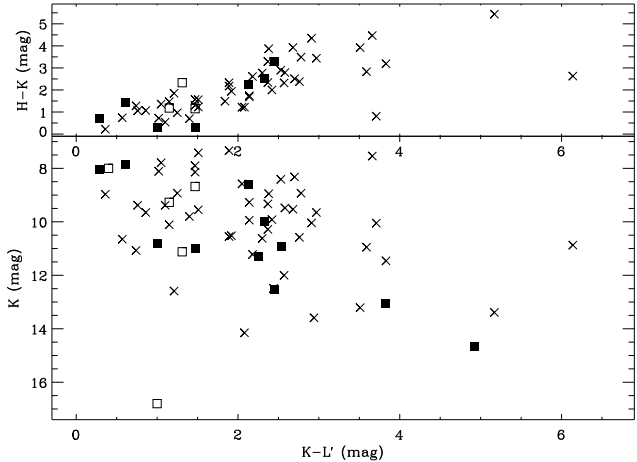}
\end{figure}

In order to get a better understanding of the nature of the companions
detected during this survey, we constructed an $H-K$/$K-L'$
color-color diagram and a $K$/$K-L'$ color-magnitude diagram. For this
purpose, we used near-IR photometry from the literature for our
sources and combined them with our relative photometry for resolved
systems. The diagrams are presented in
Figure\,\ref{fig:ccd}. Photometric information was lacking for a few
primaries in our survey, particularly in Orion, as well as for a few
companions that were not observed or detected at one or several of the
wavelengths of interest. Nonetheless, the diagrams show that the
companions lie well within the magnitude and color ranges defined by
the protostars in all clouds. The only exception is the companion to
Ser\,317\,A, which is located 5--8\,mag too low in the color-magnitude
diagram. This companion is discarded from our survey as a background
source (see Section\,\ref{sect:specials}). All other likely physical
companions appear to have colors expected for Class\,I protostars,
reinforcing the likelihood that they are related to their primaries.

The likely non-physical companions that we have identified in this
survey (see Section\,\ref{sect:specials}) are located in the same
general area of the diagrams as the primaries and singles of our
sample, though with bluer colors on average (all have $K_s-L'
<1.5$\,mag as opposed to only 3 out of 10 candidate physical
companions). This is because they are likely to be unrelated Class\,II
or Class\,III sources, which are expected to show a narrower and
not-as-red range of near-IR colors than less evolved, more embedded
objects. Their status is therefore consistent with their location in
these plots. However, it is not possible to determine the physical
nature of the companions newly identified in this survey based on
these diagrams only because the $K_s$ and $L'$ filters are too close
in wavelengths and do not probe the overall behavior of the SED of the
object in the IR. For instance, IRAS\,04248+2612, which is classified
as a Class\,I source here, has $K-L=0.57$ (Kenyon \& Hartmann 1995)
and is one of the least red objects in our sample. High-resolution
mid-IR imaging is needed to classify the new companions and study, for
instance, the pairing properties of Class\,I protostars (see Haisch et
al. 2006).

\section{Discussion}
\label{sect:discus}


\subsection{The initial rate of multiple stellar systems}

Overall, we find a high multiplicity rate among our sample of Class\,I
and FS protostars in both our Complete and Extended Surveys. To
compare these multiplicity rates to those of Main Sequence field
stars, we first note that our sample presents a wide range of
bolometric luminosities, indicating that it spans a relatively wide
range of stellar masses. With median and minimum luminosities of
$\approx2\,L_\odot$ and $\approx0.01\,L_\odot$ respectively, it is
likely that a substantial fraction of our targets will become low-mass
stars. Among field stars, the multiplicity rate of low-mass stars is
lower than that of solar-type stars (e.g., Fischer \& Marcy
1992). Directly relevant to our analysis, Delfosse et al. (2004) found
that the multiplicity rate of field M dwarfs is about 3 times lower
than that of solar-type stars in the separation range 10--1000\,AU. We
can therefore safely consider that the DM91 field star multiplicity
rate is a conservative upper limit to compare our survey to. We adopt
the Gaussian prescription for the orbital period distribution of DM91,
from which we derive a 19.5\,\% (27.8\,\%) multiplicity rate over the
same separation range of our Complete (Extended) Survey, a factor of
$\sim$1.7 lower than found in our survey. The probability of finding
18 or more companions out of 56 targets in our Complete Survey if the
intrinsic multiplicity rate were that derived from DM91 is on the
order of 3\%, i.e., this hypothesis can be rejected at the 2.2$\sigma$
confidence level. Therefore, the multiplicity excess among embedded
protostars over field stars can be considered as a robust result.

The overall multiplicity rate among Class\,I protostars is found to be
independent of the molecular cloud, within our statistical
uncertainties, even though these clouds span a wide range of physical
conditions. The Taurus and L1641 samples, with maximum stellar
densities of less than 100\,star/pc$^{-3}$ (Chen \& Tokunaga 1994;
Hartmann 2003), are representative of distributed star formation
whereas both the Ophiuchus and Serpens samples represent clustered
star formation, with densities in excess of 1000\,star/pc$^{-3}$
(Bontemps et al. 2001; Kaas et al. 2004), indicating that star
formation has indeed proceeded to different outcomes in these clouds.
Considering initial conditions, prestellar cores in Ophiuchus are on
average $\sim3$ times smaller and $\sim10$ times denser than those
found in Taurus (Motte \& Andr\'e 2001), probably a consequence of a
more violent star formation process in Ophiuchus. This could also
imply that the magnetic field is stronger in clouds like Ophiuchus and
Serpens, since its strength is correlated with the cloud local density
(Crutcher 1999). There is substantial scatter associated to this
correlation, however, and it is not yet possible to compare the
magnetic field strength in prestellar cores from one cloud to
another. Besides the L1641 YSO population, Orion\,A has also formed
the ONC, which contains high-mass stars that are not found in any of
the other three clouds studied. Because of the presence of these
high-mass stars, it is probable that the Orion\,A cloud is overall
hotter than the other clouds, although it is unclear whether this is
also true for prestellar cores, especially in the parts of L1641 that
are quite distant from the ONC. Despite these differences between
clouds, the outcome is an apparently universal multiplicity rate. This
suggests that core fragmentation proceeds to form the same proportion
of single and multiple systems irrespective of the cloud in which the
stars form. Numerical simulations and semi-analytical considerations
of cloud fragmentation suggest a strong dependence on the physical
conditions in the cloud (Bonnell et al. 1992; Durisen \& Sterzik 1994;
Boss 2002; Sterzik et al. 2003), which is not supported by our
findings. One possibility is that a self-regulatory process not
included in these simulations would lead to a similar outcome of core
fragmentation irrespective of the initial conditions. Recent
3-dimension magneto-hydrodynamics calculations suggest that the
magnetic field could play such a role (e.g., Fromang et al. 2006;
Hennebelle \& Teyssier 2007). An alternative explanation is that core
fragmentation can result in markedly different populations of multiple
systems but that early ($\lesssim10^5$\,yr) processes cancel out
initial differences although it is difficult to imagine that different
evolutionary paths lead to the same final state, however.

Since L1641 represents only a fraction of the entire Orion\,A cloud,
it may not be representative of star formation throughout the whole
molecular cloud. Unfortunately, no other survey for multiplicity among
embedded protostars has been conducted in other parts of the
cloud. Other parts of the Orion cloud, such as the ONC, host no
significant population of Class\,I sources, so that it is impossible
to determine what the initial conditions were before dynamical
evolution reshaped the original multiple systems. Nonetheless, our
survey reveals for the first time that the initial multiplicity rate
can be high even in the Orion molecular cloud, i.e., that this
property is not specific to low-mass clouds only. Furthermore,
considering the range of total masses probed by the molecular clouds
surveyed here and the size of typical star-forming regions in our
Galaxy (Adams \& Myers 2001; Lada \& Lada 2003), we argue that this
high multiplicity rate among the youngest stellar systems is the rule,
not the exception. While it is possible that much richer clusters
harbor a different proportion of multiple systems, these account for a
minor fraction of all stars in our Galaxy.


\subsection{Distribution of separations: a hint of cloud-to-cloud
  variation}

While the overall multiplicity rates appear consistent from one cloud
to another, we also searched for more subtle differences between our 4
subsamples. We find no obvious trend in the distribution of flux
ratios as a function of molecular cloud. We note, however, that it is
not possible to infer mass ratios from near-IR flux ratios for
Class\,I protostars. Rather, one needs to determine bolometric
luminosities, which require far-IR measurements, unavailable for the
companions detected in this survey.

\begin{figure}
\caption{\label{fig:histo}Distribution of projected separations for
  all companions detected in this survey and from the literature. In
  the outermost 2 bins, the solid histogram represents the
  distribution obtaining by merging this survey with the results of
  D04. The dotted histogram, which has a poorer statistical
  significance, represents the distribution from this survey
  only. Both are indistinguishable and it is therefore safe to merge
  the two surveys to improve the statistics. The vertical dashed line
  represents the innermost separation considered in our Complete
  Survey.}  \includegraphics[width=\columnwidth]{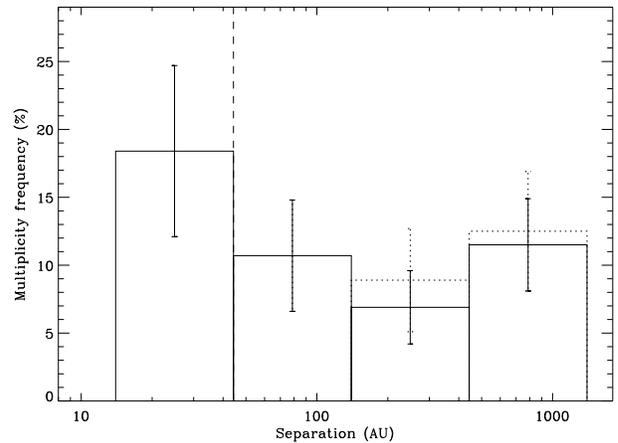}
\end{figure}

Using the distances to the molecular clouds, we transformed all
angular separations into projected physical separations. From a
statistical standpoint, the resulting distribution shown in
Figure\,\ref{fig:histo} should be similar to the distribution of
actual semi-major axes (e.g., Brandeker et al. 2006). Beyond 140\,AU,
we combined the results of this survey with that of D04 in order to
improve the statistical significance. The distribution shows a
tentative peak towards short separations, as is the case for field
solar-type binaries (DM91). The formal significance of this peak is
weak, about 1.4$\sigma$, due to our limited sample size. However, we
remind the reader that our completeness is poorer at the tightest
separations than it is for wide systems. Therefore, taking statistical
uncertainties at face value is improper in this case as there is a
systematic bias against very tight companions in our survey. Note that
if we had included the likely non-physical companions in this diagram,
a secondary peak would have been observed in the outermost bin
(19.5$\pm$4.2\% combining this survey with D04), in sharp contrast
with Main Sequence and T\,Tauri populations (DM91; Leinert et
al. 1993). While it cannot yet be excluded that such systems are
currently bound and will later be disrupted, this provides further
support to the suspicion that most of these systems are indeed
non-physical.

Interestingly, we note that the individual separation distributions
for the Taurus, Ophiuchus and Serpens clouds all cover a similar wide
range of separation and are indistinguishable. On the other hand, the
distribution in L1641 seems different: none of the companions has a
separation larger than 100\,AU (excluding companions to the high
luminosity protostars), suggesting a deficit of wider companions in
this cloud. This is confirmed in our Complete Survey by a Mann-Whitney
rank-sum statistical test which shows that the observed distribution
of separations in Orion differs from that in the 3 other clouds at the
2.8$\sigma$ level (99.7\% confidence level). While this is the most
significant difference we can identify in the properties of multiple
protostars in individual molecular clouds, the very small number of
Orion companions included here (4) warrants a more extended survey to
confirm this trend. It is nonetheless interesting to assess the
possible implication of this result.

First of all, this result could be related to the finding by Brandner
\& K\"ohler (1998) that different parts of the Upper Scorpius OB
association present different distributions of projected
separations. These authors concluded that ``same physical conditions
that facilitate the formation of massive stars also facilitate the
formation of closer binaries among low-mass stars, whereas physical
conditions unfavorable for the formation of massive stars lead to the
formation of wider binaries among low-mass stars'', a statement that
seems to apply to the Orion\,A cloud as well, if the ONC and L1641
stellar populations had similar initial multiplicity
properties. Interestingly, Sterzik et al. (2003) suggested that this
behaviour could be understood as a consequence of differences in
prestellar core temperatures: semi-analytical arguments led them to
conclude that higher core temperatures should result in tighter binary
systems. The temperature of prestellar cores in L1641 has not yet been
measured yet and there are no more prestellar cores in Upper Scorpius,
so we cannot test this scenario. The presence of high-mass stars in
these clouds suggest higher overall cloud temperatures, but this
effect may not apply to prestellar cores which are efficiently
self-shielded from surrounding ultraviolet radiation.

We also note that such a preference for very tight systems in Orion
was not found in multiplicity surveys of Class\,II objects. Previous
surveys found companions throughout the 100--1000\,AU range they
probed (Prosser et al. 1994; Padgett et al. 1997; Petr et al. 1998;
Beck et al. 2003; K\"ohler et al. 2006; Reipurth et al. 2007) even in
the ONC\footnote{Scally et al. (1999) showed that there is essentially
no binary system with a separation larger than 1000\,AU in the whole
ONC cluster, as opposed to the Taurus star-forming region or to the
field binary population, but that study focused on larger separations
than probed here.}, where prestellar core temperatures would likely
have been hottest. This seems to cast some doubt on the temperature
effect proposed by Sterzik et al. (2003), and furthermore, it is not
easy to reconcile with our findings for Class\,I protostars. If
confirmed, this result would indeed point towards a phenomenon that
repopulates the ``wide systems'' category over a timescale of
$\sim$1\,Myr. However, Class\,II surveys focused on the clustered
populations of young stars and may not be directly comparable to the
Class\,I population we have studied here. It would therefore be
premature to speculate about the nature of this pehnomenon.


\subsection{Evolution of multiple systems within molecular clouds}

If the multiplicity rate among Class\,I protostars is about 1.5 times
higher than it is in the field in all molecular clouds, as suggested
by our results, some systems must be disrupted before the populations
disperse in the interstellar medium. We analyze here a possible
scenario to account for this.

First of all, the multiplicity excess we find among Class\,I and FS
protostars is reminiscent of that found among slightly more evolved
Class\,II and III targets in Taurus and Ophiuchus (e.g., Ghez et
al. 1993; Leinert et al. 1993; Duch\^ene 1999; Ratzka et al. 2005).
Further testing the possibility of time evolution of the multiplicity
rate, we separate the Taurus and Ophiuchus targets on the basis of the
presence or absence of an extended millimeter envelope, as we have
already done in D04. No significant difference is found between the
two categories. This result seems to contradict our finding in D04
that there was a marginally significant decrease of multiplicity rate
with evolutionary status. We do not confirm this trend here, a
consequence of two factors. First, the sample sizes in the present
survey are smaller. Second, most of the companions identified here as
candidate non-physical systems (see Section\,\ref{sect:specials}) are
associated with targets possessing extended millimeter
envelopes. Indeed, re-analyzing the D04 sample excluding the candidate
non-physical systems results in a confidence level of only 82\%, down
from the original 99.6\% level. This latter point suggests that this
trend may not have as strong a statistical support as previously
thought. Overall, there appears to be no significant evolution of the
multiplicity rate as a function of evolutionary status among the YSOs
probed here. In other words, the high multiplicity rate is already set
at an age of $\sim10^5$\,yr and it hardly evolves on a timescale of
$\sim10^6$\,yr in clouds like Taurus and Ophiuchus. Comparison
surveys of samples Class\,II objects in Serpens and L1641 do not exist
yet, so we cannot extend this conclusion to the other clouds studied
here.

Ejection of companions from bound systems can have two independent
causes. First of all, if star formation proceeds to create many triple
and higher-multiplicity systems, it is possible that some of them are
unstable and eventually eject one or more of their components to reach
a stable configuration (e.g., Reipurth 2000). Alternatively, numerical
simulations have shown that wide companions can easily be stripped off
their primary if they are located in a dense stellar populations
because of the repeated near encounters of independent cluster members
(e.g., Kroupa 1995). In the following we investigate both
possibilities in the context of our observations.

The multiple systems in our survey are all triple and most of them
appear safely hierarchical, suggesting that little companions will be
expelled past the Class\,I phase. Furthermore, the ratio of binary to
triple systems in our Complete Survey is consistent with the lower
limit derived by Koresko (2002) and very similar to that obtained by
Correia et al. (2006), both of whom searched for additional companions
around known T\,Tauri binary systems. Among 52 binaries in the sample
studied by Correia et al., 7 are triple (and none quadruple) in the
45--1400\,AU separation range. The proportion of triple systems
therefore remains constant on a $\sim10^6$\,yr timescale. It is
difficult to directly compare the fraction of triple systems observed
among Class\,I protostars to that found among field stars to probe
longer timescale. As discussed in Tokovinin \& Smekhov (2002), it is
likely that the DM91 survey missed some faint, wide companions and
therefore underestimated the number of triple and higher-order systems
among field dwarfs. Correia et al. (2006) concluded that the
difference between the fraction of Class\,II and solar-type field
multiple systems is not significant. Therefore, it appears that the
disintegration of unstable multiple systems, which is frequently
observed in numerical simulations (Delgado-Donate et al. 2004; Goodwin
et al. 2004a), is limited to a timescale of less than
$10^5$\,yrs. Only future large-scale surveys of the youngest,
Class\,0, sources will eventually test this common predictions of
numerical simulations of core fragmentation. Nonetheless, the limited
number of high-order multiples found in this and other surveys,
combined with the high total multiplicity rate we have found, supports
the conclusion by Goodwin \& Kroupa (2005) that fragmentation
generally leads to systems with $N=2$ or $N=3$ components, leaving
little room for internal decays to play a role in the time evolution
of the multiplicity rate.

If companions are not ejected from unstable multiple systems, they can
be ejected as the result of the fly-by of an unrelated cloud
member. The absence of a significant decrease of the multiplicity rate
from Class\,I to Class\,II sources in a cloud like Taurus is not
surprising. The stellar density, even in the densest sub-clusters, is
too low to account for more than a handful of disrupted systems
(Kroupa \& Bouvier 2003). On longer timescale, the evolution is also
very limited, since the sub-clusters dissolve in a few Myr and each
member is then released into the galactic field, which can only
disrupt the widest systems, with separations beyond $\sim10^4$\,AU
(Weinberg et al. 1987). Therefore, the high multiplicity rate observed
in Taurus will remain almost unchanged even on a timescale of several
Gyr, implying that field stars are primarily populated from
star-forming regions where the multiplicity rate is substantially
reduced, typically denser clusters (Kroupa 1995). 

The Ophiuchus and Serpens populations can be considered as clustered
among those studied here. Indeed their peak stellar density is similar
to that of ``dominant-mode cluster'' defined by Kroupa (1995). In
Kroupa's simulations, which assume a universal multiplicity rate and
orbital period distribution, these populations will lose $\sim$40\% of
their initial companions over a timescale of $\lesssim100$\,Myr. Most
of these companions should in fact be stripped off their primaries
during the first few crossing times, which is about 3\,Myr, as the
clusters start expanding. No evolution is observed in Ophiuchus, where
Class\,II and Class\,III sources exhibit a significant multiplicity
excess (Ghez et al. 1993; Reipurth \& Zinnecker 1993; Duch\^ene
1999). Considering that the age of the stellar population is
$\lesssim1$\,Myr (Wilking et al. 1989; Greene \& Meyer 1995), i.e.,
less than the crossing time, this is not so surprizing. In Serpens,
Kaas et al. (2004) concluded that star formation underwent two
separate bursts, so that the Class\,I aggregate will probably expand
by a factor of 10 in volume over a timescale of $\sim$1\,Myr. If the
population of Class\,II sources resulting from the previous burst has
underwent such an expansion, dynamical interaction may have ripped off
some wide companions, although the cluster may not be old enough for
its multiplicity rate to have substantialy decreased
yet. Unfortunately, no high-angular resolution of Class\,II sources in
this cloud have been conducted so far.

The situation in Orion is more complex. All multiplicity surveys among
Class\,II populations in this cloud have found low multiplicity rate
in the separation range 100--1000\,AU, consistent with, or slightly
lower than, that found in field-star surveys (e.g., Padgett et
al. 1997; Petr et al. 1998; Beck et al. 2003; K\"ohler et al. 2006;
Reipurth et al. 2007). These studies have considered both the inner
and outer parts of the ONC in Orion\,A, as well as NGC\,2024, 2068 and
2071 in Orion\,B. These latter clusters are intermediate in density
between Serpens and the ONC but are all much denser than the
distributed L1641 population surveyed here. A direct comparison of
these surveys with our results is therefore not possible. However, if
companions are as frequent in all Orion sub-clusters than in L1641 on
scales of tens to hundreds of AU in the embedded phase, this implies
that many wide companions are disrupted in less than 1\,Myr. Kroupa et
al. (1999) pointed out that the ONC is probably in a rapidly expanding
phase and that it was much denser in the past. This suggests that all
stellar systems already experienced several near encounters and that
any loosely bound companion has already been ejected, even in the
outer regions. The large optical high-angular survey recently
conducted by Reipurth et al. (2007) presented supporting evidence for
this scenario. In other words, multiplicity surveys among Class\,II
and Class\,III sources are not sufficient to determine whether the
initial multiplicity rate in Orion is similar or different from other
molecular clouds (Kroupa 1995). Our finding is the first direct
evidence that the initial multiplicity rate is indeed high in Orion,
at least in some regions of the molecular cloud. Therefore, if they
can be extended to the entire Orion cloud, our results support the
view of Kroupa et al. (1999) of a binary-rich initial population that
rapidly looses most of its widest systems due to close encounters with
the cluster. In this framework, a multiplicity study of the
distributed Class\,II population of L1641, a region of moderate
stellar density, would provide the observational missing piece of the
puzzle, i.e., the possibility to compare embedded and
optically-detected YSOs in the same Orion population and to constrain
the evolution of these systems on a timescale of $\sim10^6$\,yr.


\section{Conclusion}
\label{sect:concl}

We have used diffraction-limited imaging from 1.6 to 3.7\,$\mu$m with
the 8m-VLT adaptive optics system to search for tight companions
around 45 embedded Class\,I and FS protostars in the Taurus,
Ophiuchus, Serpens and L1641 (Orion\,A) molecular clouds. We
complement our analysis with published high-resolution surveys of
similar objects in Taurus and Ophiuchus to build a sample of 58
Class\,I and FS targets. We derive an average multiplicity rate of
32$\pm$6\% over the separation 45--1400\,AU. In the Taurus and
Ophiuchus clouds, the closest clouds in our sample, we further derive
a multiplicity rate of 47$\pm$8\% over the separation range
14--1400\,AU. These rates are a factor of $\sim1.7$ higher than those
derived for nearby solar-type field stars, extending the multiplicity
excess found among several populations of T\,Tauri star to even
younger ages. Most importantly, we find that the embedded protostars
in L1641 show a multiplicity rate similar to that in other clouds,
indicating for the first time that a high multiplicity rate is
achieved after core fragmentation in all types of nearby molecular
clouds, including giant molecular clouds such as Orion\,A, which also
hosts the dense ONC cluster. We also find a high multiplicity rate in
Serpens, the densest cluster in our survey.

Our results support the view that core fragmentation results in a
multiplicity rate for wide companions that does not depend on the
initial conditions reigning in the cores, as opposed to predictions of
most numerical simulations. Rather, our findings support a scenario in
which all YSO populations start with a similar set of multiplicity
properties and only evolve as a consequence of disruptive
system-system interactions prior to dilution of the clusters in the
field. Follow-up multiplicity surveys of the Class\,II and Class\,III
populations of the Serpens and L1641 clouds would provide key
empirical verifications of this scenario. We found 6 triple systems,
all of them hierarchical, and higher-order systems are very rare among
Class\,I/FS sources within the separation ranges studied here. It is
unlikely that many companions will be ejected as a result of internal
decay of unstable systems past the Class\,I phase. Finally, we also
find a possible trend for the embedded Orion multiple systems to have
different orbital properties than those in other clouds, namely
systematically tighter projected separations, a hint that environmental
conditions may impact on the properties of protobinaries.


\begin{acknowledgements}
  We are grateful to Monika Petr-Gozens for her referee report which
  helped us improve our manuscript. We thank the ESO Paranal staff for
  their precious help during our Visiting Observing programs and ESO's
  User Support Department for our Service Observing programs. This
  research has made use of the SIMBAD database and the VizieR
  catalogue access tool, operated at CDS, Strasbourg, France, and the
  data products from the 2 Micron All Sky Survey consortium. This work
  was supported in part by the Programme National de Physique
  Stellaire of CNRS/INSU (France) and the National Science Foundation
  Science and Technology Center for Adaptive Optics managed by the
  University of California at Santa Cruz under cooperative agreement
  No. AST - 9876783.
\end{acknowledgements}


\end{document}